\documentclass{iopart}

\usepackage{epsfig}
\usepackage{amstext}
\usepackage{amssymb}

\newcommand{\ugl}{\ensuremath{U_\text{gL}}}
\newcommand{\ugr}{\ensuremath{U_\text{gR}}}
\newcommand{\ugc}{\ensuremath{U_\text{gC}}}
\newcommand{\ugx}{\ensuremath{U_\text{gX}}}
\newcommand{\ugqpc}{\ensuremath{U_\text{gQPC}}}
\newcommand{\usd}{\ensuremath{U_\text{SD}}}

\newcommand{\un}[1]{\,\text{#1}}
\newcommand{\bper}{\ensuremath{B_\perp}}

\newcommand{\goffs}{\ensuremath{G_\text{offset}}}

\newcommand{\tk}{\ensuremath{T_\text{K}}}
\newcommand{\nuzd}{\ensuremath{\nu_\text{2DES}}}
\newcommand{\gR}{\ensuremath{\mathrm g_\text{R}}}
\newcommand{\gL}{\ensuremath{\mathrm g_\text{L}}}
\newcommand{\gC}{\ensuremath{\mathrm g_\text{C}}}
\newcommand{\gX}{\ensuremath{\mathrm g_\text{X}}}
\newcommand{\gQPC}{\ensuremath{\mathrm g_\text{QPC}}}

\begin{document}

\title{A widely tunable few electron droplet}

\author{A~K~H\"uttel$^1$\footnote{Present address: Molecular Electronics and
    Devices, Kavli Institute of Nanoscience, Delft University of Technology,
    PO Box 5046, 2600 GA Delft, The Netherlands}, K~Eberl$^2$\footnote{Present
    address: Lumics GmbH, Carl--Scheele--Strasse 16, 12489 Berlin, Germany}
    and S~Ludwig$^1$}
\address{$^1$~Center for NanoScience and Department f\"ur Physik,
Ludwig--Maximilians--Universit\"at, Geschwister--Scholl--Platz 1,
80539~M\"unchen, Germany}\ead{A.K.Huettel@tudelft.nl}
\address{$^2$~Max-Planck-Institut f\"ur Festk\"orperforschung,
Heisenbergstra{\ss}e 1, 70569 Stuttgart, Germany}

\begin{abstract}
Quasi--static transport measurements are employed to characterize a few electron
quantum dot electrostatically defined in a GaAs/AlGaAs heterostructure. The gate
geometry allows observations on one and the same electron droplet within a wide
range of coupling strengths to the leads. The weak coupling regime is described
by discrete quantum states. At strong interaction with the leads Kondo phenomena
are observed as a function of a magnetic field. By varying gate voltages the
electron droplet can, in addition, be distorted into a double quantum dot with a
strong interdot tunnel coupling while keeping track of the number of trapped
electrons.
\end{abstract}


\maketitle

\section{Introduction}

Extensive 
experimental work has recently been aimed towards electrostatically
defining and controlling semiconductor quantum dots~\cite{kouwenhoven-report-1,prb-ciorga:16315,prb-elzerman:161308,prl-petta:186802,prb-pioro-ladriere:125307}.
These efforts are impelled by proposals
for using localized electron spin~\cite{pra-loss:120} or charge
states~\cite{jjap-vanderwiel:2100}, respectively, as qubits, the elementary
registers of the hypothetical quantum computer. 
The complete control of the
quantum dot charge down to the limit of only one trapped
conduction band electron was demonstrated by monitoring single electron tunneling (SET) current
through the device as well as by a nearby charge
detector~\cite{prb-ciorga:16315, prl-field:1311, prl-sprinzak:176805}.

In this article, we present data on an electron droplet in which the charge can
be controlled all the way to the limit of one electron. The quantum dot is
defined electrostatically by using split gates on top of an epitaxially grown
AlGaAs/GaAs heterostructure. We observe a wide tunability of the electronic
transport properties of our device. Recent work
focused either on the case of weak coupling between a quantum dot and its
leads~\cite{prb-ciorga:16315}, or on the Kondo regime of strong coupling to the
leads~\cite{prl-sprinzak:176805}. Here, we explore a structure that can be
fully tuned between these limits. In addition, we demonstrate how the shape of
the quantum dot confinement potential can be
distorted within the given gate geometry \cite{prb-kyriakidis:035320} all the way into a double well
potential describing a double quantum dot~\cite{anticrossing,ep2ds,kondo}. The charge
of the electron droplet can be monitored during the deformation process.

The heterostructure used for the measurements embeds a two-dimensional
electron system (2DES) $120\un{nm}$ below the crystal surface. The electron
sheet density and mobility in the 2DES at the temperature of $T=4.2\un{K}$ are
$n_\text{s} \simeq 1.8\times 10^{15}\,\text{m}^{-2}$ and $\mu \simeq 75
\,\text{m}^2/\text{Vs}$, respectively. We estimate the 2DES temperature
to be of the order $T_\text{2DES} \sim 100\,\text{mK}$.

Our gate electrode geometry for defining a quantum dot, shown in the SEM micrograph of 
%
%%%%%%%%%%%%%%%%%%%%%%%%%%%%%%%%%%%%%%%%%%%%%%%%%%%%%%%%%%%%%%%%%%
% PLACE PICTURE HERE FOR 4-pageversion
\begin{figure}[tb]
\begin{center}
\epsfig{file=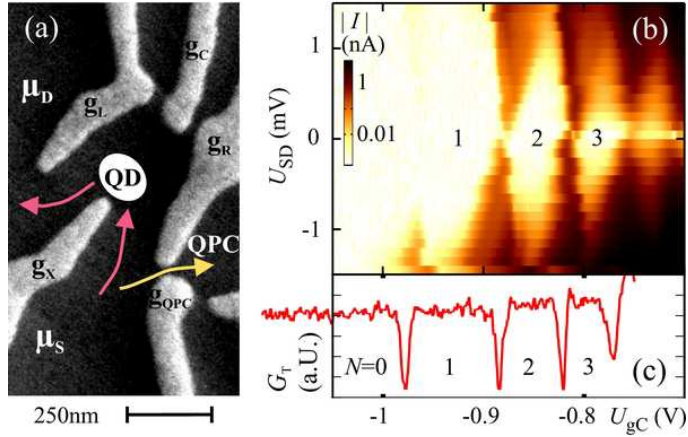, width=9cm}
\end{center}
\vspace*{-0.4cm}
\caption{
  (Color online) (a) SEM micrograph of the gate electrodes used to
  electrostatically define a quantum dot (marked as QD) and a quantum point
  contact (marked as QPC). (b) Exemplary measurement of the absolute value of 
  the SET current $I$ through the quantum dot as a
  function of the center gate voltage $\ugc$ and the bias voltage $\usd$. (c)
  Differential transconductance $G_\text T(\ugc)$ of the QPC measured at
  identical parameters as in (b) but for $\usd=0$. The numerals $N=0,\,1,\,2,\,3$
  in (b) and (c) depict the actual number of conduction band electrons trapped
  in the quantum dot.
}
\label{fig1}
\end{figure}
%%%%%%%%%%%%%%%%%%%%%%%%%%%%%%%%%%%%%%%%%%%%%%%%%%%%%%%%%%%%%%%%%%
%
Fig.~\ref{fig1}(a), is designed following a geometry introduced by Ciorga 
{\it et al.}~\cite{prb-ciorga:16315}. Because of the triangular shape of
the confinement potential, an increasingly negative voltage on the plunger gate
\gC\ depletes the quantum dot and simultaneously shifts the potential minimum
towards the tunnel barriers defined by gates \gX\ and \gL, or \gX\ and \gR,
respectively. This way, the tunnel barriers between the leads and the
electron droplet can be kept transparent enough to allow the detection of SET
current through the quantum dot even for an arbitrarily small number of trapped
conduction band electrons~\cite{prb-ciorga:16315}.

Fig.~\ref{fig1}(b) shows an exemplary color scale plot of the
measured quantum dot SET current $\left| I \right|$ as a function of the gate voltage \ugc\ and
the source drain voltage \usd. Within the diamond-shaped light regions in
Fig.~\ref{fig1}(b) SET is hindered by
Coulomb blockade and the charge of the quantum dot is constant. The gates marked
\gR\ and \gQPC\ in Fig.~\ref{fig1}(a) are used to define a quantum point contact
(QPC). As demonstrated in Refs.~\cite{prl-field:1311} and
\cite{prl-sprinzak:176805}, a nearby QPC can provide a non-invasive way to
detect the charge of a quantum dot electron by electron. The result of such a
measurement is shown in Fig.~\ref{fig1}(c), where  the transconductance $G_\text
T=\text{d}I_\text{QPC} / \text{d}\ugc$ obtained using a lock-in amplifier is plotted for $\usd\simeq 0$, along the
corresponding horizontal trace in Fig.~\ref{fig1}(b). Note that
Figs.~\ref{fig1}(b) and (c) have identical $x$ axes.
The advantage of using a QPC charge
detector is that its sensitivity is almost independent of the quantum dot charge
state. In contrast, the current through the quantum dot decreases as it is discharged
electron by electron, because of an increase of the tunnel barriers between the
quantum dot and the leads. This can be clearly seen by a comparison of the
magnitude of the current oscillations in Fig.~\ref{fig1}(b) with the
transconductance minima in Fig.~\ref{fig1}(c).\footnote{An apparent double
  peak structure in Fig.~\ref{fig1}(b) around $\usd \sim 0$ can be explained
  by noise rectification effects.} The QPC
transconductance measurement plotted in Fig.~\ref{fig1}(c) shows no pronounced
local minima corresponding to changes of the quantum dot charge for $\ugc <
-1\un{V}$. This indicates that the quantum dot is here entirely
uncharged. This observation has been confirmed by further careful tests as
e.g.\ variation of the tunnel barriers or variation of the QPC lock-in
frequency and QPC bias. The inferred number of conduction band
electrons $N=0,\,1\,,\dots$ trapped in the quantum dot is indicated in
the Coulomb blockade regions in Figs.~\ref{fig1}(b) and
\ref{fig1}(c). \footnote{The SET current shown in Fig.~\ref{fig1}(b) between
  $N=0$ and 
  $N=1$ can not be resolved for $\usd\sim 0$. We ascribe this to an asymmetric
  coupling of the quantum dot to the leads.} 

In the following we demonstrate the flexibility provided by the use of voltage
tunable top-gates for a lateral confinement of a 2DES. We first focus on the
regime of a few electron quantum dot weakly coupled to its leads, where the
shell structure of an artificial two-dimensional atom in the circularly
symmetric case is described by the
Fock--Darwin states~\cite{zphys-fock:446,proccam-darwin:86}. Secondly, we present measurements with the quantum dot
strongly coupled to its leads. Here we observe Kondo features. Finally, we
explore the deformation of the few electron droplet into a serial double
quantum dot by means of changing gate voltages. The transport spectrum of this
artificial molecule has been described in previous publications for the
low electron number limit ($0\le N\le 2$)~\cite{anticrossing,ep2ds,kondo,mauterndorf}.

\section{Weak coupling to the leads}

\begin{figure}[th]\begin{center}
\epsfig{file=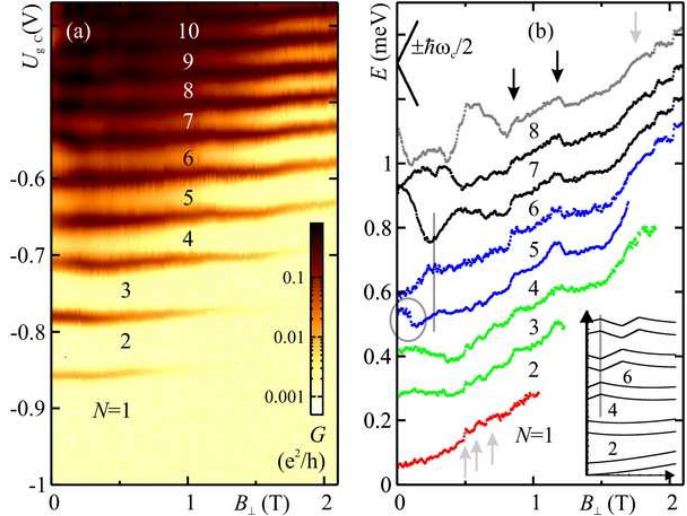, width=9cm}
\end{center}
\vspace*{-0.4cm}
\caption{ 
(Color online) (a) Differential conductance $G$ of the quantum dot in dependence
on a magnetic field \bper\ perpendicular to the 2DES and the voltage on gate
\gC. All other gate voltages are kept fixed (see main text). 
(b) \bper-field dependence of a relative energy corresponding to the local
maxima of G. The traces are numerically obtained from the measurement
shown in (a) after a conversion of the gate voltage to energy and subtraction
of an arbitrary but \bper-field independent
energy, respectively. Black arrows mark common features of all traces.
A gray vertical line indicates the first ground state transition of the quantum
dot for $N \gtrsim 4$. Inset: Qualitative prediction for the traces, using a
Fock-Darwin potential and the constant interaction model.
}
\label{fig2}
\end{figure}

The regime of a few electron quantum dot weakly coupled to its leads is reached
for gate voltages of $\ugl=-0.52\un{V}$, $\ugr=-0.565\un{V}$, and
$\ugx=-0.3\un{V}$. The observed  Coulomb blockade oscillations are shown in
Fig.~\ref{fig2}(a), 
%
%%%%%%%%%%%%%%%%%%%%%%%%%%%%%%%%%%%%%%%%%%%%%%%%%%%%%%%%%%%%%%%%%%
% PLACE PICTURE HERE FOR 4-pageversion
%%%%%%%%%%%%%%%%%%%%%%%%%%%%%%%%%%%%%%%%%%%%%%%%%%%%%%%%%%%%%%%%%%
%
where the differential conductance $G\equiv\text{d}I/\text{d}\usd$ of the quantum dot is
plotted in a logarithmic (color) scale as a function of center gate voltage
\ugc\ and magnetic field perpendicular to the 2DES \bper. The absolute
number $N$ of trapped electrons within the Coulomb blockade regions,
derived by means of the QPC charge detection, is indicated
by numerals.

The characteristic \bper-field dependence of the local maxima of differential
conductance in Fig.~\ref{fig2}(a), marking the Coulomb oscillations of SET, has
also been observed via capacitance spectroscopy of lateral quantum
dots~\cite{prl-ashoori:613} and via transport spectroscopy of vertically etched
quantum dots~\cite{prl-tarucha:3613}.

The addition energy of a quantum dot for each electron number $N$ can be derived from the
vertical distance (in \ugc) between the local SET maxima,
by converting the gate voltage scale \ugc\ into a local potential energy. The
conversion factor for the present quantum dot has been obtained from nonlinear transport
measurements; a constant conversion factor is used as first-order
approximation~\cite{kouwenhoven-report-1}. Accordingly, 
in Fig.~\ref{fig2}(b) the \bper\ dependence of the differential conductance
maxima positions is plotted after conversion to energy scale. The traces are obtained by
numerically tracking the local SET maxima in Fig.~\ref{fig2}(a). An arbitrary but
\bper-independent energy is subtracted from each trace, such that all
traces are equidistant at $\bper=1\un{T}$ -- i.e.\ at a magnetic field high
enough such that orbital effects are not relevant to the \bper\ dependence of the addition enery
anymore. For a direct comparison the inset of 
Fig.~\ref{fig2}(b) displays the \bper-dependence expected within the so-called constant
interaction model~\cite{kouwenhoven-report-1}, that approximates many particle
effects with a classical capacitance term, for the so-called Fock-Darwin states. 
These are solutions of the single particle Schr\"odinger equation of a ``two-dimensional atom''.
In detail the vector potential of \bper\ and the Fock-Darwin potential
$V= m^\ast\omega_0^2 r^2/2$ are considered. The latter describes a
two-dimensional harmonic oscillator with
characteristic frequency $\omega_0$ and effective electron mass $m^\ast$, at the
distance $r$ from its potential minimum~\cite{zphys-fock:446,proccam-darwin:86}. The
harmonic approximation is justified for a few electron quantum dot with a
relatively smooth electrostatic confinement as usually provided by remote gate
electrodes.

Although not necessarily expected for lateral quantum dots, where the tunnel
barriers to the 2DES leads automatically induce symmetry breaking, 
for electron numbers $1 \le N\le 7$ the measured \bper\ dependence
(Fig.~\ref{fig2}(b)) resembles these model expectations (inset). The observed and
predicted pairing of SET differential conductance maxima corresponds to an
alternating filling of two-fold spin-degenerate levels~\cite{prl-tarucha:3613,
nature-fuhrer:822, prl-luscher:2118}.

A local maximum of addition energy is visible at $N=6$, which would correspond
to a filled shell in a circular symmetric potential~\cite{prl-tarucha:3613}.
For $4\le N\le 7$ the first orbital ground
state transition is visible as cusps at $0.25\un{T} \lesssim \bper \lesssim
0.3\un{T}$. The cusps are marked by a vertical gray line in Fig.~\ref{fig2}(b) and its
inset, respectively. The magnetic field at which this
transition happens allows to estimate the characteristic energy scale of the
confinement potential~\cite{rpp-kouwenhoven:701}
$\hbar \omega_0= \sqrt{2} \, \hbar \omega_c(\bper) \sim 680\,\mu\text{eV}$.
The expected maximum slopes of the $E(\bper)$ traces are given by the
orbital energy shift and expected to be in the order of $\text d E/\text d \bper
= \pm \hbar\omega_c /2B$, where $\omega_c = e \bper/m^\ast$ is the cyclotron
frequency in GaAs. These expected maximum slopes are indicated in the upper left
corner of Fig.~\ref{fig2}(b) and agree well with our observations.

For the $4\le N\le 5$ transition and at a small magnetic field $\bper \lesssim 
0.2\un{T}$ our data exhibit a pronounced cusp marking a slope reversal, as indicated
by a gray ellipsoid in Fig.~\ref{fig2}(b). Assuming a circularly symmetric
potential, this deviation from the
prediction within the constant interaction model can be understood in terms of
Hund`s rules by taking into account the exchange coupling of two electron
spins~\cite{prl-tarucha:3613}. Along this model the exchange energy can be
estimated to be $J\sim 90\,\mu\text{eV}$ for the involved states. Interestingly,
an according deviation from the constant interaction model for the $3\le N \le
4$ transition~\cite{prl-tarucha:3613} predicted by Hund's rules is not observed
in our measurement. This behavior might be related to a possibly more asymmetric
confinement potential at lower electron number, lifting the required orbital level degeneracy.
For $N\ge 7$ the $E(\bper)$ traces do not anymore resemble the
Fock-Darwin state predictions. We attribute this to modifications of the transport spectrum
caused by electron-electron interactions. In addition, the measurements 
plotted in Fig.~\ref{fig2}(a) indicate strong co-tunneling currents within
the Coulomb blockade regions for $N\gtrsim 7$. This can be seen by the
growing conductance in the Coulomb blockade regions as the electron
number is increased.

At the magnetic fields of $\bper \simeq 0.88\,\text{T}$ and $\bper \simeq
1.17\,\text{T}$ all traces exhibit a common shift, as marked by black arrows
in Fig.~\ref{fig2}(b). This may be explained by an abrupt change of the
chemical potential in the leads, since at these magnetic fields the 2DES in
the leads reaches even integer filling factors of $\nuzd=8$ and $\nuzd=6$,
respectively.\footnote{A step-like feature in the data at $\bper \simeq 1.75\un{T}$ can be identified 
with the filling factor $\nuzd=4$ (gray arrow in Fig.~\ref{fig2}(b)), 
however here the observation is far less clear
than at $\nuzd=6$ and $\nuzd=8$. 
At higher filling factors $\nuzd=10, 12, \dots$ (also gray arrows) the effect 
diminishes and is 
partially shadowed by the orbital transitions.}  The integer filling factors of the 2DES have been identified in
the Coulomb blockade measurements up to $\nuzd=1$ at $\bper \simeq 7.1\un{T}$,
where as in previous publications~\cite{prb-ciorga:16315} also a shift at odd
\nuzd\ is observed (data not shown).

\section{Strong coupling to the leads}

By increasing the voltages on the side gates \ugl\ and \ugr\ the quantum dot in the
few electron limit is tuned into a regime of strong coupling to the leads.
During this process the position of the SET differential conductance maxima is
tracked so that the quantum dot charge state remains well known. At strong
coupling we observe enhanced differential conductance in Coulomb blockade
regions due to the Kondo effect~\cite{ptp-kondo:37, nature-goldhaber-gordon:156,
prl-goldhaber-gordon:5225}.

\begin{figure}[th]
\begin{center}
\epsfig{file=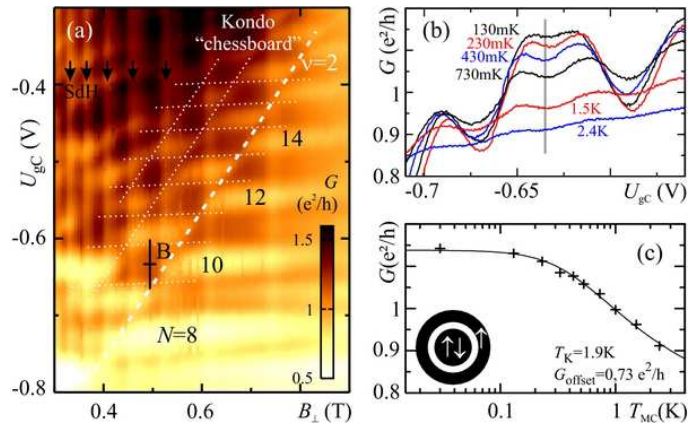, width=9cm} 
\end{center} 
\vspace*{-0.4cm} 
\caption{
(Color online) (a) Differential conductance $G$ at strong coupling to the leads
as a function of perpendicular magnetic field \bper\ and gate voltage \ugc. A
distinct chessboard-like pattern of enhanced conductance is observed (see dotted
lines). Black arrows mark Shubnikov-de-Haas conductance minima of the 2DES in the leads. (b)
Conductance traces $G(\ugc)$ at constant $\bper=495\un{mT}$ for different
cryostat temperatures. The traces are measured along the vertical line marked with ``B''
in (a). (c) Cryostat temperature dependence of the conductance $G$ at $\bper=495\un{mT}$
and $\ugc=-0.635\un{V}$ (vertical gray line in (b)). The solid line is a model
curve for a Kondo temperature of $T_\text{K}=1.9\un{K}$ (see text for
details).
}
\label{fig3}
\end{figure}

Fig.~\ref{fig3}(a)
%
%%%%%%%%%%%%%%%%%%%%%%%%%%%%%%%%%%%%%%%%%%%%%%%%%%%%%%%%%%%%%%%%%% 
% PLACE PICTURE HERE FOR 4-pageversion 
%%%%%%%%%%%%%%%%%%%%%%%%%%%%%%%%%%%%%%%%%%%%%%%%%%%%%%%%%%%%%%%%%%
%
shows part of the transport spectrum of the quantum dot as a
function of \bper\ and \ugc\ at $\ugl=-0.508\un{V}$,
$\ugr=-0.495\un{V}$, and $\ugx=-0.3\un{V}$. Compared to the weak coupling case
displayed in Fig.~\ref{fig2} the SET differential conductance maxima (almost
horizontal lines) are broader in Fig.~\ref{fig3}. This broadening can be
explained by a much stronger coupling to the leads.
In addition, a background differential conductance increases
monotonously towards more positive gate voltage \ugc. This background is
independent of the Coulomb blockade oscillations. The quantum dot is here
near the mixed valence regime where charge quantization within the
confinement potential is lost. Thus, the conductance background is explained by
direct scattering of electrons across the quantum dot. Vertical lines of decreased
differential conductance, marked in Fig.~\ref{fig3}(a) with black arrows,
indicate minima in the density of states at the Fermi energy of the lead 2DES
caused by Shubnikov-de-Haas oscillations.

Between the maxima of SET differential conductance Coulomb blockade is
expected. Instead we observe a distinct chessboard-like
pattern of areas of enhanced or supressed differential conductance, in the
region
highlighted by the white dashed or dotted lines in Fig.~\ref{fig3}(a). This
feature is independent of the Shubnikov-de-Haas oscillations (vertical lines).
Similar phenomena have already been observed in many-electron quantum dots and
have been identified as a \bper-dependent Kondo effect~\cite{prl-schmid:5824}. The magnetic
field perpendicular to the 2DES leads to the formation of Landau-like core and
ring states in the quantum dot, as sketched in
Fig.~\ref{fig3}(c)~\cite{prl-sprinzak:176805, prb-keller:033302}. The electrons
occupying the lowermost Landau level effectively form an outer ring and dominate
the coupling of the quantum dot to its leads, whereas the higher Landau-like
levels form a nearly isolated electron state in the core of the quantum
dot~\cite{prb-mceuen:11419, prl-vaart:320, prl-stopa:046601}. On one hand, with
increasing magnetic field one electron after 
the other moves from the core into the outer ring, and hence the total spin of the
strongly coupled outer ring can oscillate between $S=0$ and $S=1/2$. Only for
a finite spin the Kondo-effect causes an enhanced differential conductance. On the
other hand a change in \ugc\ eventually results in a change of the total number
and total spin of the conduction band electrons trapped in the quantum dot. 

In addition, charge redistributions between the Landau-like levels of the quantum
dot may influence the SET maxima positions~\cite{prl-sprinzak:176805, prb-mceuen:11419,
prl-stopa:046601}. The combination
of these effects explains the observed chessboard-like pattern of enhanced and
supressed differential conductance through the quantum dot. 
For a higher magnetic field where the filling factor falls below $\nu=2$ inside the
electron droplet the separation in outer ring and core state does not exist
anymore. The chessboard-like 
pattern disappears and the Kondo-effect is expected to depend monotonously on
\bper. Indeed, for \bper\ larger than a field marked by the dashed white line in
Fig.~\ref{fig3}(a) the Kondo-current stops to oscillate as a function of \bper.
From this we conclude that the dashed white line in
Fig.~\ref{fig3}(a) identifies the $\nu=2$ transition inside the quantum dot.

Fig.~\ref{fig3}(b) displays exemplary traces $G(\ugc)$ of the differential
conductance as a function of the gate voltage \ugc\ at a fixed magnetic field
$\bper=495\un{mT}$ for different cryostat temperatures. These traces are taken
along the black vertical line in Fig.~\ref{fig3}(a) marked by `B`. The vertical
line in Fig.~\ref{fig3}(b) marks the expected position of a minimum of the
differential conductance due to Coulomb blockade, as indeed observed for the
traces recorded at high temperature. At low temperature, instead of a
minimum an enhanced differential conductance is measured due to the Kondo
effect. Note, that the two minima of the differential conductance adjacent to
the Kondo feature in
Fig.~\ref{fig3}(b) show the usual temperature behavior indicating the here the
Kondo effect is absent (in accordance with the chessboard-like pattern in
Fig.~\ref{fig3}(a)). Fig.~\ref{fig3}(c) displays the differential conductance at
the center of the Coulomb blockade region marked by the vertical line in
Fig.~\ref{fig3}(b), as a function of the cryostat temperature. The solid line is
a model curve given by $G(T)= G_0 \left(
{T_\text{K}'^2}/ \left( T^2 + T_\text{K}'^2 \right) \right)^{s} + \goffs$ with
$T_\text{K}'=\tk /{\sqrt{2^{1/s}-1}}$~\cite{prl-goldhaber-gordon:5225}. The low temperature limit of the
Kondo differential conductance $G_0$ is taken as a free parameter, as well as an offset
$\goffs$ that has been introduced to take into account the effect of the
temperature-independent background current described above. For $s=0.22$ as
expected for spin-$1/2$ Kondo effect~\cite{prl-goldhaber-gordon:5225} we find
best agreement between the model and our data at a Kondo temperature of $\tk
=1.9\un{K}$, a limit Kondo conductance $G_0=0.41 \, e^2/h$ and a conductance offset
$\goffs=0.73 \, e^2/h$.  All nearby areas of enhanced Kondo differential conductance
display a similar behaviour with Kondo temperatures in the range of $1.2\un{K}
\lesssim \tk \lesssim 2.0\un{K}$. 

In addition, the dependence of the differential conductance $G$ on the
source-drain voltage \usd\ has been measured for different regions of the
parameter range in Fig.~\ref{fig3}(a) (data not shown). These measurements are 
fully consistent with above results. They display a zero-bias
conductance anomaly in the high conductance 'Kondo' regions, that can be
suppressed by changing the magnetic field \bper.

\section{Deformation into a double quantum dot}

\begin{figure}[tb]\begin{center}
\epsfig{file=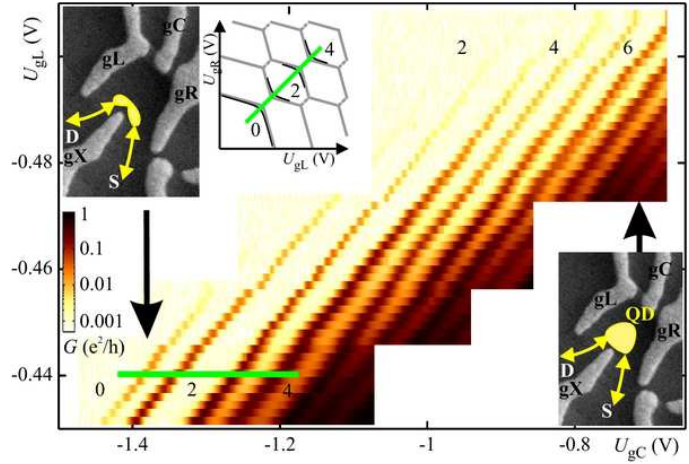, width=9cm}
\end{center}
\vspace*{-0.4cm}
\caption{ 
  (Color online) Differential conductance of the electron droplet as a function of \ugc\
  (x axis) and the simultaneously varied side gate voltages $\ugl \propto \ugr$
  (y axis). As the gate voltage is decreased below $\ugc \simeq -1.2\un{V}$
  lines of conductance maxima form pairs with smaller distance, indicating the
  deformation of the quantum dot into a double quantum dot (see text). Insets: A
  SEM micrograph of the top gates with sketches of the approximate potential
  shapes of the quantum dot or double quantum dot. The third inset shows a
  sketch of the stability diagram as expected for the case of a double quantum
  dot. The thick solid lines are guides for the eye.}
\label{fig4}
\end{figure}

The shape of the confinement potential of our quantum
dot can be modified by changing the voltages applied to the
split gate electrodes. This is a general feature of electrostatically defined
structures in a 2DES. A non-parabolic confinement potential is e.g.\ discussed
by the authors of Ref.~\cite{prb-kyriakidis:035320}. Here, we demonstrate a controlled
deformation of the confinement potential, transforming one local minimum, i.e.\ a quantum
dot, into a double well potential describing a double quantum dot. 
Such a transition is shown in Fig.~\ref{fig4},
%
%%%%%%%%%%%%%%%%%%%%%%%%%%%%%%%%%%%%%%%%%%%%%%%%%%%%%%%%%%%%%%%%%%
% PLACE PICTURE HERE FOR 4-pageversion
%%%%%%%%%%%%%%%%%%%%%%%%%%%%%%%%%%%%%%%%%%%%%%%%%%%%%%%%%%%%%%%%%%
%
which plots Coulomb blockade oscillations of differential conductance (color scale) in
dependence of the center gate voltage \ugc\ along the x-axis. We aim to
transform a quantum dot charged by $N=0,\,1,\, 2,\,...$ electrons into a
peanut-shaped double quantum dot with the same charge
(see insets of Fig.~\ref{fig4}). This is done by creating a high
potential ridge between gates \gX\ and \gC , i.e.\ by making \ugc\ more
negative. In order to keep the overall charge of our device constant, both side
gate voltages \ugl\ and \ugr\ (y-axis) are changed in the opposite direction
than \ugc. For the opposed center gate \gX\ we choose $\ugx=-0.566\un{V}$,
causing a significantly higher potential than in the previous measurements. 

For $\ugc\gtrsim -1\un{V}$ the Coulomb oscillations are in first order
quasiperiodic, as can be seen in the upper right quarter of
Fig.~\ref{fig4}. 
This is expected for a single quantum dot with addition energies large compared
to the orbital quantization energies.
In contrast, for more negative \ugc\ an onset of 
a doubly periodic behavior is observed. I.e.\ along the thick 
solid horizontal line in the lower left corner of Fig.~\ref{fig4} the 
distance between adjacant conductance maxima oscillates, most clearly visible for $N<4$.
Such a doubly periodic behaviour is expected for a double quantum dot in case of a symmetric
double well potential. This is the case along the thick solid line in
the inset of Fig.~\ref{fig4} sketching the double quantum dot's stability
diagram. In a simplified picture, if the double
quantum dot is charged by an odd number of electrons the charging energy for the
next electron is approximately given by the int{\sl er}dot Coulomb repulsion of
two electrons separated by the tunnel barrier between the adjacent quantum
dots.  However, for an even number of electrons the charging energy for the next
electron corresponds to the larger int{\sl ra}dot Coulomb repulsion between two
electrons confined within the same quantum dot. Therefore, the difference
between int{\sl er}dot and int{\sl ra}dot Coulomb repulsion on a double
quantum dot causes the observed doubly periodic oscillation. 

The asymmetry of the double quantum dot with respect to the potential minima of
the double well potential can be controlled by means of the side gate voltages
\ugl\ and \ugr . Coulomb blockade results in a stability diagram characteristic
for a double quantum dot as sketched in an inset of Fig.~\ref{fig4} in
dependence of the side gate voltages~\cite{prb-hofmann:13872, prl-blick:4032,
rmp-wiel:1}. Gray lines separate areas of stable charge configurations. The
corners where three different stable charge configurations coexist are called
triple points of the stability diagram. For a serial double quantum dot with
weak interdot tunnel coupling, the charge of both quantum dots can fluctuate
only near the triple points and only here current is expected to flow. The
bisector of the stability digram (solid bold line in the inset) defines a
symmetry axis, along which the double well potential and, hence, the charge
distribution in the double quantum dot is symmetric.  In the case of two (one)
trapped conduction band electrons we identify our structure as an artificial
two-dimensional helium (hydrogen) atom that can be continuously transformed into
an (ionized) molecule consisting of two hydrogen atoms. 

\begin{figure}[tb]\begin{center}
\epsfig{file=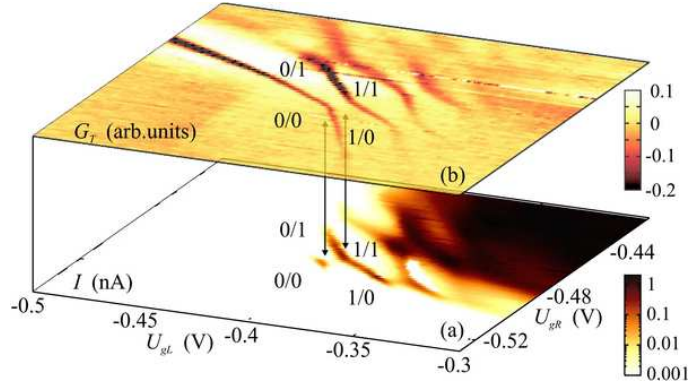, width=9cm}
\end{center}
\vspace*{-0.4cm}
\caption{ 
(Color online) (a) Dc-current through the double quantum dot, (b)
transconductance $G_\text{T} \equiv \text{d}I_\text{QPC}/\text{d}U_\text{gL}$ of the nearby QPC
used as a double quantum dot charge sensor, with identical axes \ugl\ and
\ugr. The additional gate voltages are in both plots chosen as $\ugc=-1.4\un{V}$, 
$\ugx=-0.566\un{V}$, and $\ugqpc=-0.458\un{V}$.
}
\label{fig5}
\end{figure}

To prove the presence of a few electron double quantum dot after performing the
described transition, we plot in Fig.~\ref{fig5} 
%
%%%%%%%%%%%%%%%%%%%%%%%%%%%%%%%%%%%%%%%%%%%%%%%%%%%%%%%%%%%%%%%%%%
% PLACE PICTURE HERE FOR 4-pageversion
%%%%%%%%%%%%%%%%%%%%%%%%%%%%%%%%%%%%%%%%%%%%%%%%%%%%%%%%%%%%%%%%%%
%
the measured stability diagram of our device. Fig.~\ref{fig5}(a) shows the linear
response dc current through the device ($\usd=50\,\mu\text{V}$) as a function of the side gate voltages
\ugl\ and \ugr. Fig.~\ref{fig5}(b) displays the QPC
transconductance $G_\text{T}\equiv \text{d}I_\text{QPC}/\text{d}U_\text{gL}$. The areas of stable charge configurations are
marked by numerals indicating the number of conduction band electrons in the
left / right quantum dot~\cite{anticrossing, ep2ds}. Both plots clearly feature
areas of stable charge configurations separated by either a current maximum (in
(a)) or a transconductance minimum (in (b)), respectively. 
The transconductance measurement confirms the electron numbers obtained from
the single QD case, as even for very asymmetric confinement potential no
further discharging events towards more negative gate voltages \ugl\ and \ugr\ are observed.
In comparison to
the gray lines in the inset of Fig.~\ref{fig4} the edges of the hexagon pattern are here
strongly rounded. This indicates a sizable interdot tunnel coupling that cannot be
neglected compared to the interdot Coulomb interaction~\cite{anticrossing,
ep2ds}. A large interdot tunnel coupling results in molecular states delocalized
within the double quantum dot. This additionally explains the observation of finite current
not only on the triple points of the stability diagram, but also along edges of
stable charge configurations in Fig.~\ref{fig5}(a). Here the total charge
of the molecule fluctuates, allowing current via a delocalized state. In
previous publications the low-energy spectrum of the observed double well potential was
analyzed and the tunability of the tunnel coupling demonstrated~\cite{anticrossing, ep2ds}.

\section*{Summary}

Using a triangular gate geometry, a highly versatile few electron quantum dot has been
defined in the 2DES of a GaAs/AlGaAs heterostructure. The couplings between the
quantum dot and its leads can be tuned in a wide range. For weak quantum dot --
lead coupling, the shell structure of the states for $1 \lesssim N \lesssim 7$
trapped conduction band electrons is observed. The transport spectrum supports
the assumption of a Fock-Darwin like trapping potential and subsequent filling
of spin-degenerate states. A deviation from the model prediction can be
partially explained by the alignment of spins according to Hund's rule. For
strong quantum dot -- lead coupling, a chessboard pattern of regions of
enhanced zero bias conductance in dependence of a magnetic field perpendicular
to the 2DES is observed. The enhanced conductance regions are explained in terms
of the Kondo effect, induced by the formation of Landau-like core and ring
states in the quantum dot. Finally, for strongly negative center gate voltages,
the quantum dot trapping potential can be distorted at constant charge into a
peanut shaped double quantum dot with strong interdot tunnel coupling.

\section*{Acknowledgements}

We like to thank L. Borda and J.\,P.\ Kotthaus for helpful discussions. We
acknowledge financial support by the Deutsche Forschungs\-ge\-mein\-schaft via
the SFB 631 ``Solid state based quantum information processing'' and the
Bundesministerium f\"ur Bildung und Forschung via DIP-H.2.1. A.\, K.\ H\"uttel
thanks the Stiftung Maximilianeum for support.

%%%%%%%%%%%%%%%%%%%%%%%%%%%%%%%%%%%%%%%%%%%%%%%%%%%%%%%%%%%%%%%%%%%%
%% BIBLIOGRAPHY

\section*{References}

\bibliographystyle{iopart-num}

\bibliography{paper,diss}

\end{document}